\documentclass[aps,prl,reprint,showpacs,superscriptaddress,amsmath,amssymb]{revtex4-1}

\usepackage{graphicx}
\usepackage{bm}
\usepackage[colorlinks=true]{hyperref}
\usepackage{color}

\newcommand{\Tc}{T_{\rm c}} 
 
\newcommand{\tU}{\tilde{U}} 
\newcommand{\tUmin}{\tilde{U}_{\rm min}} 
\newcommand{\kmin}{k_{\rm min}} 

\begin{document}

\title{Controlled Self-Assembly of Periodic and Aperiodic Cluster Crystals}

\author{Kobi Barkan}
\affiliation{Raymond and Beverly Sackler School of Physics and Astronomy,
Tel Aviv University, Tel Aviv 69978, Israel}
\author{Michael Engel}
\affiliation{Department of Chemical Engineering, University of Michigan, Ann Arbor, MI 48109, USA}
\author{Ron Lifshitz}
\email[Corresponding author:\ ]{ronlif@tau.ac.il}
\affiliation{Raymond and Beverly Sackler School of Physics and Astronomy,
Tel Aviv University, Tel Aviv 69978, Israel}
\affiliation{Condensed Matter Physics 149-33, California Institute of
  Technology, Pasadena, CA 91125, USA}
\date{January 16, 2014}

\begin{abstract}
  Soft particles are known to overlap and form stable clusters that self-assemble into periodic crystalline phases with density-independent lattice constants.  We use molecular dynamics simulations in two dimensions to demonstrate that, through a judicious design of an isotropic pair potential, one can control the ordering of the clusters and generate a variety of phases, including decagonal and dodecagonal quasicrystals.  Our results confirm analytical predictions based on a mean-field approximation, providing insight into the stabilization of quasicrystals in soft macromolecular systems, and suggesting a practical approach for their controlled self-assembly in laboratory realizations using synthesized soft-matter particles.
 \end{abstract}


\maketitle

Particles interacting via pair potentials with repulsive cores, which are either bounded or only slowly diverging---like those found naturally in soft matter systems~\cite{Likos01a}---can be made to overlap under pressure to form clusters~\cite{Klein94,*Likos98}, which then self-assemble to form crystalline phases~\cite{Stillinger76,*Likos02, *Mladek06, *Zhang10}. The existence of such cluster crystals was recently confirmed in amphiphilic dendritic macromolecules using monomer-resolved simulations~\cite{Lenz12}, and in certain bosonic systems~\cite{Cinti10, *Henkel12}. They occur even when the particles are purely repulsive, and typically exhibit periodic fcc or bcc structures. Here we employ molecular dynamics (MD) simulations in two dimensions, guided by analytical insight, to show how isotropic pair potentials can be designed to control the self-assembly of the clusters, suggesting a practical approach that could be applied in the laboratory. We obtain novel phases, including a striped (lamellar) phase and a hexagonal superstructure, as well as decagonal (10-fold) and dodecagonal (12-fold) quasicrystals.

Given a system of $N$ particles in a box of volume $V$, interacting via an isotropic pair potential $U(r)$ with a repulsive core, a sufficient condition for the formation of a cluster crystal is a negative global minimum $\tUmin=\tU(\kmin)<0$ in the Fourier transform of the potential~\cite{Likos01b,*Likos07}.  This condition implicitly requires the potential not to diverge too strongly, so that the Fourier transform exists.  The wavenumber $\kmin$ determines the length scale for the order in the system by setting the typical distance between neighboring clusters.  Above a sufficiently high mean particle density $\bar{c}=N/V$, a further increase of $\bar{c}$ increases the number of overlapping particles within each cluster, but does not change the distance between their centers. It also determines the critical temperature $k_{\mathrm{B}}\Tc=-\tUmin \bar{c}$,~\cite{Likos01b,*Mladek06, *Likos07, *Mladek07} below which the liquid becomes unstable against crystallization, where $k_{\mathrm{B}}$ is the Boltzmann constant.  

\begin{figure*}
\centering
\includegraphics[width=0.95\textwidth]{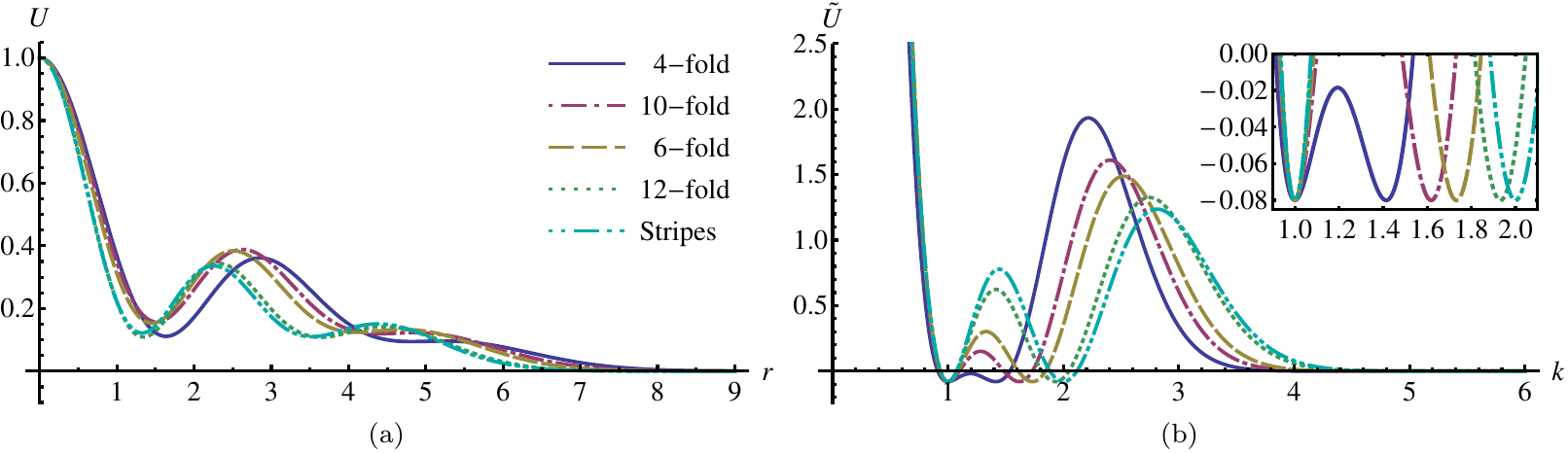}
\caption{(Color online) \label{fig:Pair-potentials}
Pair potentials used in this study. (a) Real space $U(r)$, normalized such that $U(0)=1$. (b)~Fourier transform $\tilde{U}(k)$, where the first minimum is always at $k=1$, and the second minimum is located with increasing order at $k_4=\sqrt{2}$ (solid blue), $k_5=(1+\sqrt5)/2$ (dot-dashed red), $k_6=\sqrt3$ (dashed brown), $k_{12}=\sqrt{2+\sqrt3}$ (dotted green), and $k_\infty=2$ (double-dot-dashed cyan). The inset shows a close-up view of the minima.}
\end{figure*}

As the particles form increasingly larger clusters, the system becomes well characterized by a continuous coarse-grained density function $c(\bm{r})$.  The thermodynamic behavior can then be described in a mean-field approximation, which becomes exact in the high-density high-temperature limit~\cite{Lang00, *Louis00}. Using MD simulations, we examine the analytical predictions of a particular mean-field approximation that was proposed by Barkan, Diamant, and Lifshitz (BDL)~\cite{Barkan11} to explain the stability of a certain class of soft quasicrystals in two dimensions.  BDL confirmed an earlier conjecture~\cite{LD07} that attributed the stability of soft quasicrystals to the existence of two length scales in the pair potential, combined with effective many-body interactions arising from entropy. Accordingly, we study pair potentials whose Fourier transforms contain two negative minima of the same depth $\tUmin$, like the ones shown in Fig.~\ref{fig:Pair-potentials}. That stable quasicrystals may need two length scales in their effective interaction potentials is not new~\cite{Olami90, *Smith91}.  Many two-length-scale potentials were investigated over the years and found to exhibit stable quasiperiodic phases~\cite{Dzugutov93, *Jagla98, *Skibinsky99, *Quandt99, *Roth00, *Keys07, *Engel07, *Engel08a, *Molinero10, *Archer13}. The novelty of BDL was in their quantitative understanding of the stabilization mechanism, allowing them to pinpoint regions of stability in the parameter spaces of different potentials, instead of performing an exhaustive search. The inclusion of a second length scale provides greater control over the self-assembly of the clusters than what can be achieved with a single minimum only. We demonstrate how this enables one to generate a wide range of novel periodic and aperiodic cluster crystal structures.  

Our findings come at a time when an ever-growing number of soft-matter systems are found to exhibit phases with quasiperiodic long-range order---all showing dodecagonal symmetry~\cite{ungar05,*ungar11,*Dotera11,*Dotera12}.  First discovered in liquid crystals made of amphiphilic dendritic macromolecules~\cite{Zeng04}, self-assembled soft quasicrystals have since appeared in ABC-star polymers~\cite{takano05,*hayashida07}, in systems of nanoparticles~\cite{talapin09,*bodnarchuk13}, with hard tetrahedra~\cite{Haji-Akbari09}, in block co-polymer micelles~\cite{Fischer11,*Iacovella11}, and in mesoporous silica~\cite{Xiao12}.  These systems provide exciting platforms for the fundamental study of the physics of quasicrystals~\cite{LifshitzIJC11} and promise new applications of self-assembled nanomaterials~\cite{electronics,*selfassemblyreview}.

The key idea of BDL was borrowed from a model developed by Lifshitz and Petrich (LP)~\cite{LP97}, who extended the Swift-Hohenberg equation~\cite{swift77} to study parametrically excited surface waves (Faraday waves), also exhibiting dodecagonal quasiperiodic order~\cite{edwards93}.  The Swift-Hohenberg equation is a generic model for pattern-forming systems~\cite{Cross09} that describes the instability of a uniform state against the formation of Fourier modes with a fixed and finite wavenumber.  In the LP modification the instability occurs simultaneously at two wavenumbers, whose ratio $q$ is tunable.  It is then the role of resonant three-mode interactions to stabilize structures containing triplets of Fourier modes with wave vectors that add up to zero. By setting the value of the wavenumber ratio $q$ to $k_n \equiv 2\cos(\pi/n)$, one can form triplets containing two unit wave-vectors separated by $2\pi/n$, and a third wave vector of length $k_n$.  Indeed, stable patterns with $n$-fold symmetry were shown to exist in the LP model for $n=4$, 6, and 12, with wavenumber ratios $k_4=\sqrt2$, $k_6=\sqrt3$, and $k_{12}=\sqrt{2+\sqrt3}$, respectively, as well as stripes for $k_\infty=2$~\cite{LP97}.  Patterns with 8-fold symmetry are unstable within the LP model, but there is a narrow window of stability for 10-fold patterns with a ratio of $k_{5}=(1+\sqrt5)/2$, although not with $k_{10}=\sqrt{(5+\sqrt5)/2}$~\cite{BL-inprep}.

Based on these design principles, and to remain as pedagogical as possible, we work directly in Fourier space to construct the family of smooth pair potentials shown in Fig.~\ref{fig:Pair-potentials}. Yet, we emphasize that our approach can be applied to any realistic potential with sufficiently many tunable parameters. We use a polynomial in even powers of the wavenumber $k$, such that two equal-depth minima can explicitly be positioned at 1 and $q=k_n$, similar in form to the effective potential used by LP. We then multiply this polynomial by a Gaussian to limit the extent of the potential. These LP-Gaussian potentials are given in Fourier space by \begin{equation}
  \label{eq:PairPotentialsFourierSpace}
  \tU(k) = e^{-\frac{k^2}{2\sigma^2}} \left(D_0 + D_2 k^2 + D_4 k^4 + D_6 k^6 + D_8 k^8\right),
\end{equation}
and are self-dual in the sense that they have the same functional form in real space. Using a two-dimensional Fourier transform we obtain
\begin{align}
  \nonumber
  U(r) &= \frac{1}{2\pi}\int_0^\infty \tU(k) J_0(kr) k\, {\rm d}k\\
  &= e^{-\frac12 \sigma^2 r^2} \left(C_0 + C_2 r^2 + C_4 r^4 + C_6 r^6 + C_8 r^8\right),
  \label{eq:PairPotentialsRealSpace}
\end{align}
where $J_0$ is the zeroth-order Bessel function, and the polynomial coefficients $C_0,\ldots,C_8$ are linear functions of the Fourier-space coefficients $D_0,\ldots,D_8$. We set the latter five independent coefficients such that $U(0)=C_0=1$, and there are two equal-depth minima at positions 1 and $k_n$ in reciprocal space with $\tU(1)=\tU(k_n)=-0.08$~\cite{Supp}. This sets both the energy scale and the length scale in our description of the problem.  The standard deviation $\sigma$ of the Gaussian in reciprocal space is chosen such that the potentials are purely repulsive in real space, although this is not required.  Note that while it is difficult to tell the potentials apart in real space (Fig.~\ref{fig:Pair-potentials}(a)) and therefore not obvious to anticipate which cluster crystal they will stabilize, the potentials are clearly distinguishable in reciprocal space (Fig.~\ref{fig:Pair-potentials}(b)), where the wavenumber ratio $q$ is visible. Similarly, one would need to tune the real-space parameters of any realistic potential to possess the required minima in Fourier space. The LP-Gaussian potentials benefit from being simple, bounded, and rapidly decaying and therefore amenable to MD simulations.

\begin{figure*}
\includegraphics[width=1.0\textwidth]{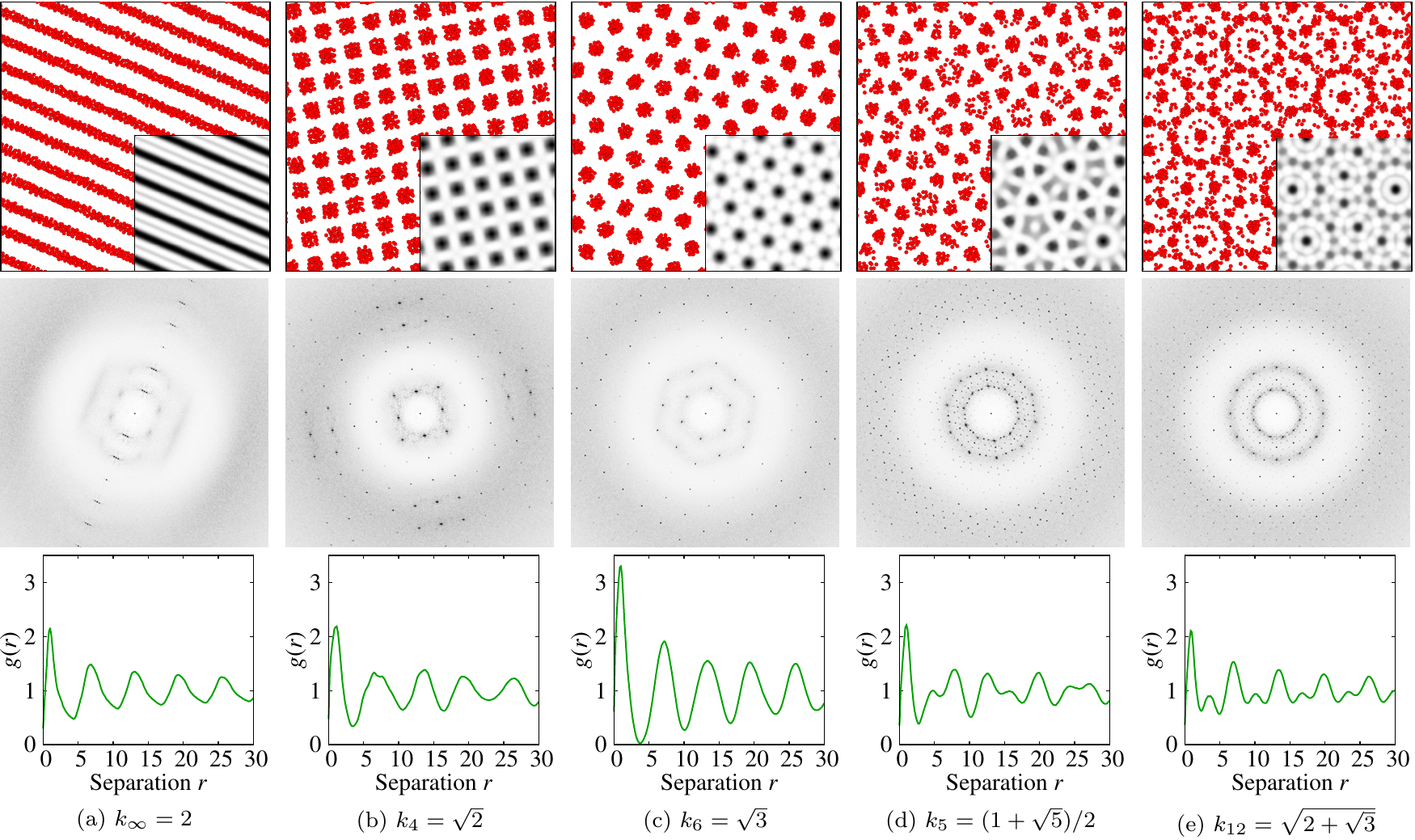}
\caption{(Color online)
  Self-assembled cluster crystals for the pair potentials in Fig.~\ref{fig:Pair-potentials}.
  We generate (a)~stripes, periodic (b)~tetragonal and (c)~hexagonal, as well as quasiperiodic (d)~decagonal and (e)~dodecagonal cluster crystals. Simulation parameters are $N=16384$, $T=0.03$ and $\bar{c}=0.8$~(a,e), $\bar{c}=0.9$~(b), $\bar{c}=0.7$~(c), $\bar{c}=0.6$~(d). The top, middle, and bottom rows show snapshots in real space comparing MD results (red circles) with mean-field predictions for $c(\bm{r})$ (in grayscale at the bottom-right corners), diffraction diagrams, and radial distribution functions. Particles are drawn with radius 1. The real-space view is limited to about 20\% of the simulation box~\cite{Supp}.
  \label{fig:Molecular-dynamics-simulation}
}
\end{figure*}

We initialize GPU-accelerated MD simulations~\cite{[{}][{. HOOMD-blue web page: \url{http://codeblue.umich.edu/hoomd-blue}.}]Anderson08} in the liquid phase above the melting temperature. The system is slowly cooled down in the $NVT$ ensemble to induce self-assembly. At $T=0$, the protocol is reversed until melting occurs. Typically, the first signs of ordering are strong density fluctuations in the liquid, which then condense into clusters and spread to develop global order.  Individual particles can migrate between neighboring clusters at elevated temperatures, even after cluster crystallization has set in, to average out density fluctuations and heal defects.  We observe self-assembly at all densities in the studied range $0.1\le\bar{c}\le2.0$.  While at low densities, $\bar{c}\le0.5$, the particles behave more individually and the hexagonal crystal prevails, the equilibrium patterns at higher densities follow the predictions of mean-field theory. We find striped (lamellar), tetragonal, hexagonal, decagonal, and dodecagonal cluster crystals (Fig.~\ref{fig:Molecular-dynamics-simulation}). In all cases, the strongest peaks in the diffraction diagrams are located at $1$ and $k_n$, the two minima of $\tU(k)$, followed by a ring with little scattering where $\tU(k)$ has its maximum. The superstructure of secondary lamellae and clusters in the striped and hexagonal phases, respectively, predicted by the mean-field densities in Fig.~\ref{fig:Molecular-dynamics-simulation}(a,c), is observed in the MD simulations at higher densities~\cite{Supp}. The latter is analogous to the hexagonal super\-lattice structures observed in Faraday-wave experiments~\cite{Arbell02}.

\begin{figure}
\centering
\includegraphics[width=0.9\columnwidth]{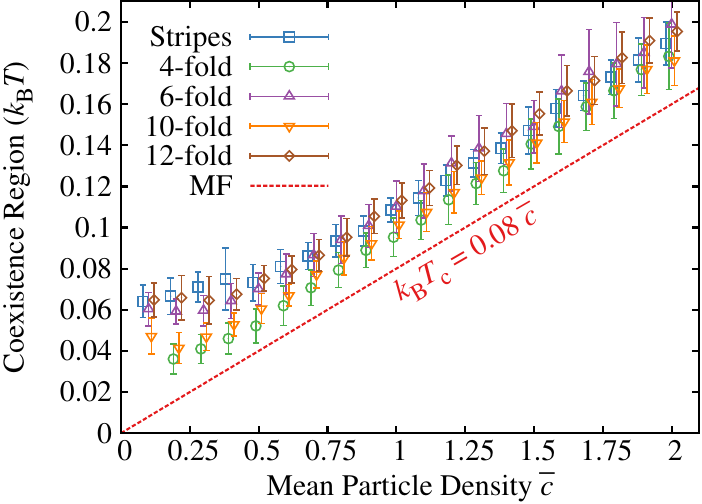}
\caption{(Color online)
  Coexistence of the liquid and cluster crystal phases, shown as bars, as a function of mean particle density $\bar{c}$ from simulation data. The dashed red line is the mean-field (MF) instability limit. Densities are slightly shifted horizontally by $\pm0.01$ among different cluster crystals for better visibility.
  \label{fig:transitionTemp}
}
\end{figure}

The transition from the liquid to a cluster crystal is a first order phase transition and therefore accompanied by hysteresis. We use ``error'' bars in Fig.~\ref{fig:transitionTemp} to show the temperature range of coexistence, obtained from simulation, as a function of $\bar{c}$.  The bars span the temperature range from where crystallization is observed upon cooling to where melting occurs upon heating. The lower ends are bounded from below by the mean-field predicted $\Tc$, shown in Fig.~\ref{fig:transitionTemp} as a dashed straight line. Except for low densities, where the mean-field approximation fails, we observe a shift of about $0.01$-$0.03$ between the simulations and the mean-field line. This shift is due to thermal fluctuations causing the cooled liquid to become unstable earlier.

\begin{figure}
\includegraphics[width=1.0\columnwidth]{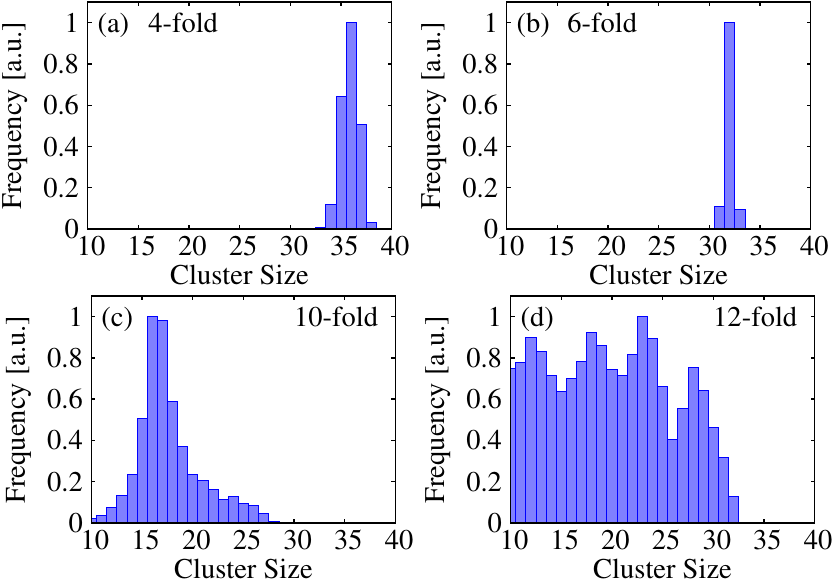}
\caption{(Color online)
  Histograms of the cluster sizes in Fig.~\ref{fig:Molecular-dynamics-simulation}.
  (a,b)~Periodic crystals show sharp distributions.
  (c,d)~Quasicrystals exhibit broad distributions.
  We use the cluster size cutoff parameter $\text{MinPts}=8$ for the DBSCAN algorithm~\cite{Ester96}.
  \label{fig:Histogram} }
\end{figure}

We further characterize the ordered phases by identifying individual clusters using the DBSCAN algorithm~\cite{Ester96}.
Fig.~\ref{fig:Histogram} demonstrates that while the cluster size distribution is narrowly peaked for periodic cluster crystals indicating a single characteristic cluster size, the cluster size distribution has a broad peak for the decagonal cluster crystal and is flat and almost featureless for the dodecagonal cluster crystal.  This observation is in agreement with experimentally observed distributions of high-symmetry stars in quasiperiodic light fields~\cite{Mikhael10} and with the mean-field density profiles shown alongside the MD simulation results in Fig.~\ref{fig:Molecular-dynamics-simulation}.

\begin{figure}
\centering
\includegraphics[width=1.0\columnwidth]{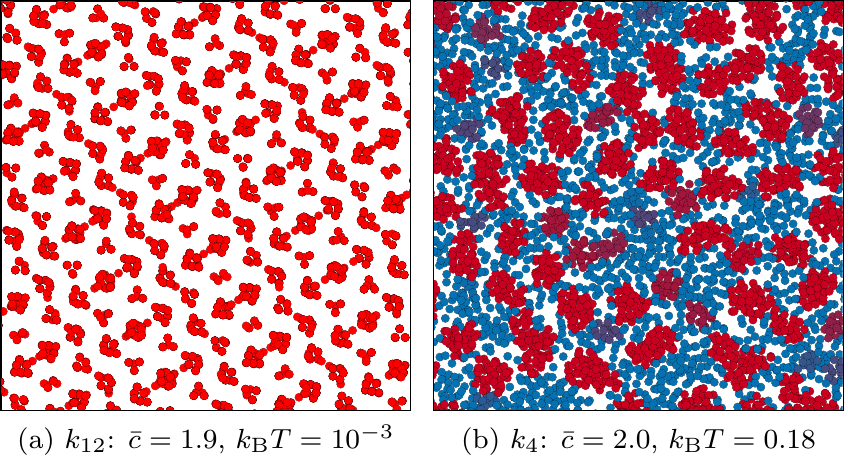}
\caption{(Color online)
  (a)~At low temperature and high density the dodecagonal quasicrystal can reversibly transform into a compressed hexagonal phase~\cite{Supp}. (b)~The liquid phase right above the onset of ordering, possibly showing a liquid of clusters. Clusters are colored according to their size from small (blue, size $\le10$) to large (red, size $\ge30$).
  \label{fig:Snapshots}
}
\end{figure}

At lower temperatures, mean-field theory predicts that the quasicrystals should become unstable toward a secondary transformation into a periodic phase of lower rotational symmetry, such as a hexagonal cluster crystal~\cite{Barkan11}. We do not observe a transformation for the decagonal quasicrystal. We do observe a secondary transformation for the dodecagonal quasicrystal into either the so-called $\sigma$ phase~\cite{Supp}, which is a known periodic approximant for dodecagonal quasicrystals that is commonly observed in soft-matter systems~\cite{Zeng04,takano05,talapin09,Lee10}, or a compressed hexagonal phase (Fig.~\ref{fig:Snapshots}(a)), similar to the one considered by LP in their Fig.~2(b).  In all cases we find that the transformation is reversible and the quasicrystal re-forms in simulation upon heating, confirming the mean-field prediction regarding the role of entropy in its stabilization.  We frequently observe significant density fluctuations already prior to ordering, as shown in Fig.~\ref{fig:Snapshots}(b), indicating the possibility that first a cluster liquid is formed, and then the clusters order. This opens up interesting questions about the formation mechanism of the cluster crystals. The observation that cluster crystals sometimes `lock-in' their orientation to the simulation box~\footnote{We confirmed the occurrence of the lock-in phenomena, where a main crystallographic axis is oriented parallel to the simulation box, in about $10$-$20\%$ of the simulations with $N=16384$ using two independent MD codes.} suggests that fluctuations are important and classical nucleation theory might not be applicable.

To conclude, we have shown how to control the self-assembly of a variety of cluster crystals by using isotropic pair potentials with two length scales, and designing their ratio in Fourier space---a general procedure that can be applied to other kinds of potentials and in the lab.  This work can be continued in several directions.  Longer and larger simulations are necessary to accurately identify the stability regions and better characterize the cluster crystals that form.  The dynamics leading to crystallization and the study of collective phonon and phason degrees of freedom in the ordered state are open problems.  Finally, an extension to three dimensions is a next step toward making a firmer contact with experimental observations of quasicrystals in soft matter systems.

We are grateful to Michael Cross, Haim Diamant, Jens Glaser, Liron Korkidi, and Goran Ungar for fruitful discussions. This research is supported by the Israel Science Foundation through Grant No.~556/10.

\bibliography{ClusterCrystals}

\end{document}